\documentclass[preprint2, trackchanges, twocolumn]{aastex701}

\usepackage{mathtools}
\usepackage{todonotes}
\usepackage{xcolor}

\def\code#1{\texttt{#1}}

\defcitealias{EHTM87Paper5}{EHTC~M87*~V}
\defcitealias{EHTSgrAPaper5}{EHTC~Sgr~A*~V}

\begin{document}
\title{\texttt{Jipole}: A Differentiable \texttt{ipole}-based Code for Radiative Transfer in Curved Spacetimes}

\author[orcid=0000-0002-9318-9329]{Pedro Naethe Motta}
\email[show]{pedronaethemotta@usp.br}
\affiliation{Instituto de Astronomia, Geof\'{\i}sica e Ci\^encias Atmosf\'ericas, Universidade de S\~ao Paulo, S\~ao Paulo, SP 05508-090, Brazil.}
\affiliation{Computational Physics and Methods (CCS-2), Center for Nonlinear Studies (CNLS) \& Center for Theoretical Astrophysics (CTA), Los Alamos National Laboratory, Los Alamos NM 87545, USA}
\correspondingauthor{Pedro Naethe Motta}

\author[orcid=0000-0002-0393-7734]{Ben S. Prather}
\affiliation{Computational Physics and Methods (CCS-2), Center for Nonlinear Studies (CNLS) \& Center for Theoretical Astrophysics (CTA), Los Alamos National Laboratory, Los Alamos NM 87545, USA}
\affiliation{Black Hole Initiative at Harvard University, 20 Garden Street, Cambridge, MA 02138, USA}
\email[]{bprather@fas.harvard.edu}

\author[orcid=0000-0001-9528-1826]{Alejandro C\'ardenas-Avenda\~no}
\affiliation{Computational Physics and Methods (CCS-2), Center for Nonlinear Studies (CNLS) \& Center for Theoretical Astrophysics (CTA), Los Alamos National Laboratory, Los Alamos NM 87545, USA}
\affiliation{Department of Physics, Wake Forest University, Winston-Salem, North Carolina 27109, USA}
\email[]{cardenas@wfu.edu}

\begin{abstract}
Recent imaging of supermassive black holes by the Event Horizon Telescope (EHT) has relied on exhaustive parameter-space searches, matching observations to large, precomputed libraries of theoretical models. As observational data become increasingly precise, the limitations of this computationally expensive approach grow more acute, creating a pressing need for more efficient methods. In this work, we present \code{Jipole}, an automatically differentiable (AD), \texttt{ipole}-based code for radiative transfer in curved spacetimes, designed to compute image gradients with respect to underlying model parameters. These gradients quantify how parameter changes—such as the black hole’s spin or the observer’s inclination—affect the image, enabling more efficient parameter estimation and reducing the number of required images. We validate \code{Jipole} against \code{ipole} in two analytical tests and then compare pixelwise intensity derivatives from AD with those from finite-difference methods. We then demonstrate the utility of these gradients by performing parameter recovery for an analytical model in three increasingly complex cases for the injected image: ideal, blurred, and blurred with added noise. In most cases, high-accuracy fits are obtained in only a few optimization steps, failing only in cases with extremely low signal-to-noise ratios. These results highlight the potential of AD-based methods to accelerate robust, high-fidelity model-data comparisons in current and future black hole imaging efforts.
\end{abstract}

\section{Introduction} 

Current approaches for interpreting event-horizon scale images of supermassive black holes rely on extensive forward modeling of emission, usually from general-relativistic magnetohydrodynamic (GRMHD) simulations~\citep{Wong:2022rqr}.  For example, the Event Horizon Telescope Collaboration (EHTC) used a forward-modeling approach with GRMHD to interpret their total-intensity images of the black hole event horizon regions of M87* and Sgr~A* (\citealt{EHTM87Paper5,EHTSgrAPaper5}, hereafter \citetalias{EHTM87Paper5,EHTSgrAPaper5}), which have provided insight into black hole spin, disk magnetization, and viewing angle.

The libraries of images required for testing forward models can contain millions of simulated snapshots (60,000 images in \citetalias{EHTM87Paper5}, 5.5 million in \citetalias{EHTSgrAPaper5}), which are stored and individually compared to measurements in order to evaluate models. Despite the existence of several GPU-accelerated ray-traced radiative transfer codes, such as \code{GRay}~\citep{Chan_2013}, \code{Odyssey}~\citep{Pu_2016}, \code{ipole}~\citep{Monika_2018, Moscibrodzka_2023}, or \code{RAPTOR}~\citep{Bronzwaer_2018}, the overall workflow can face bottlenecks in storing and reading the large quantities of GRMHD snapshots and image data involved. These constraints make libraries cumbersome to produce, store, and analyze.

The computational challenges of image libraries will only become more pressing with the advent of next-generation instruments. Planned upgrades to the Event Horizon Telescope ground-based array, such as the Next Generation Event Horizon Telescope (ngEHT)~\citep{Johnson:2023ynn}, and proposed space–very long baseline interferometry (VLBI) missions, such as the Black Hole Explorer (BHEX)~\citep{Johnson:2024ttr}, will deliver higher angular resolution and dynamic range. The better accuracy will allow new tests of strong-field gravity and deeper insights into the plasma environment surrounding supermassive black holes like M87* and Sgr A*, but will make full, forward-modeled image libraries even larger and more costly.

By focusing on posterior inference of image parameters directly from observational data, rather than exhaustively forward-modeling large libraries of simulations, one can reduce or even eliminate the need for precomputed image libraries. This strategy enables a broader and more flexible exploration of parameter space, including, for example, variations in electron temperature prescriptions or alternative particle energy distribution models. Within the EHTC, parameter estimation is already performed using Markov Chain Monte Carlo (MCMC) methods, which have been successfully applied to both emissivity-based and GRMHD image models \citep{Broderick_2020, Yfantis_2024}. Building on this foundation, more sophisticated MCMC analyses have incorporated semianalytical descriptions and additional model complexities \citep{Palumbo_2022, Tiede_2022, Yfantis_2024_bipole, Keeble_2025}. However, MCMC becomes computationally demanding when gradients of the likelihood function are unavailable. To address this, recent work has introduced differentiable analytical ray-tracing models via automatic differentiation (AD) \citep{Chang_2024, Sharma:2023nbk}. In this work, we advance these ideas by extending AD to the geodesic and radiative transfer equations, thereby enabling direct gradient-based inference of physical parameters such as the observer’s inclination and the black hole spin. The same framework can easily accommodate differentiation with respect to the plasma physical parameters, such as angular momentum, magnetic field or electron heating. This implementation will be explored in future work.

\setcounter{footnote}{0}

To address some of these these computational bottlenecks, storage demands, and scalability challenges, we present \code{Jipole}\footnote{The code used for these simulations is publicly available at \url{https://github.com/pedronaethe/Jipole} and is preserved on Zenodo~\cite{naethe_motta_zenodo_2025_17353576}.}, an imaging tool written in \code{julia} and built upon the general relativistic ray-tracing framework of \code{ipole} \citep{Monika_2018}. Using the AD implementation in the \textit{ForwardDiff.jl} package~\citep{RevelsLubinPapamarkou2016}, \code{Jipole} can compute intensity differentials with respect to input parameters. This allows us to fit synthetic images to observational data using gradient-aware optimization methods, improving both the convergence speed and the precision of the fitting process. In contrast to \citet{Sharma:2023nbk}, who used AD to compute Christoffel symbols during transport to increase speed and accuracy, we produce a fully differentiable result, including exact differentials of every pixel intensity against the fitting parameters. The examples presented in this work should be regarded primarily as a technical exercise and proof of concept. While our differentiable framework demonstrates the feasibility and potential advantages of gradient-based fitting for simplified analytical models, applying it directly to real observational data would require a significantly more robust and comprehensive modeling framework, which we leave for future development.

The rest of this paper is organized as follows: Section~\ref{sec:governing_equations} presents the governing equations solved by \code{Jipole}. Section~\ref{sec:numerical_methods} details the integration schemes employed, describes how intensity differentials are propagated through the ray-tracing integration, and outlines the simple conjugate gradient fitting procedure used for testing. In Section~\ref{sec:validation}, we validate \code{Jipole} against \code{ipole} in two distinct cases: a thin-disk analytical model~\citep{Prather_2023} and an emission-absorption model~\citep{Gold_2020}, also comparing intensity differentials obtained via AD and finite differences (FD). Section~\ref{sec:results} presents fitting examples, using a simple CG scheme to recover parameters of an analytical model under three scenarios: ideal (no blurring or noise), blurred, and blurred with noise. Lastly, Section~\ref{Sec:Conclusions} summarizes our findings and discusses potential future developments.

\section{Governing Equations} \label{sec:governing_equations}

In this section, we briefly describe the main ingredients for solving the radiative transfer equation. We refer the reader to~\cite{Wong:2022rqr} for further details.

Our main goal is to compute the intensity observed by a distant observer generated by an accretion disk around a Kerr black hole. The path the generated photons take is described by the geodesic equation
\begin{equation}
    \frac{d^2x^\mu}{d\lambda^2} = \Gamma^\mu_{\alpha \beta} k^\alpha k^\beta,
    \label{eq:sec-order_geo}
\end{equation}
where $x^\mu$ and $k^\mu$ are the four-position and four-momentum of the photon, respectively, $\lambda$ is the affine parameter of the geodesic, $\Gamma^\mu_{\alpha\beta}$ are the connection coefficients, and we have used Einstein summation notation to write these expressions. These second order equations can be divided into the following set of two first order differential equations
\begin{gather}
     \frac{dx^\mu}{d\lambda} = k^\mu ,
    \label{eq:pos_geo}
    \\
    \frac{dk^\mu}{d\lambda} = \Gamma^\mu_{\alpha \beta} k^\alpha k^\beta,
    \label{eq:mom_geo}
\end{gather}
which will be explicitly used when solving the radiative transfer equation. For computational efficiency, the geodesics are traced backwards from the camera to the black hole. Once the geodesics are computed, we forward integrate the intensity of the light for each pixel on the camera following the covariant description of the radiative transfer equation
\begin{equation}
    \frac{d}{d\lambda}\left(\frac{I_\nu}{\nu^3}\right) = \left( \frac{j_{\nu}}{\nu^2}\right) - (\alpha_\nu \nu) \left(\frac{I_\nu}{\nu^3} \right),
    \label{eq:rad_trans_inv}
\end{equation}
where $\nu$ is the photon's frequency, $I_\nu$ is the specific intensity, $j_\nu$ is the specific emissivity and $\alpha_\nu$ is the absorption coefficient. The quantities enclosed in Eq.~\ref{eq:rad_trans_inv} by parentheses are relativistically invariant quantities. For the simulations performed in this work, $j_{\nu}$ and $\alpha_{\nu}$ are simplified, idealized functions of the photon's 4-position and 4-momentum, intended to emulate emission and absorption from the plasma. The emission and absorption coefficients can be easily modified to accommodate other radiative processes without the need to change the underlying framework of the code. The specific description for $j_{\nu}$ and $\alpha_{\nu}$ adopted for the tests in this work is presented in detail in  Sections~\ref{sec:thin_disk_comparison} and \ref{sec:analytic_model_comparison}. We will now describe how we will solve Eq.~\ref{eq:rad_trans_inv} and include the information about the parameters that the specific intensity depends on.

\subsection{Derivative Evolution}

We are interested in the evolution of the intensity with regard to each input parameter $P$, or the intensity differentials ($d I_{\nu} / d P$). In this work, as a first step, we will consider both the observer's inclination angle ($\theta_o$) and the black hole dimensionless spin ($a$) as elements of the parameter set $P$. In general, this parameter set can be quite large, as it can also contain the astrophysical parameters of the emitting source. Assuming that the equations are well-behaved (continuous and differentiable), we use the formalism described in~\cite{Willkom_2018} to write the derivative, with respect to the parameters, of the relevant equations described above as
\begin{gather}
    \frac{d}{d\lambda} \frac{dx^\mu}{dP} = \frac{dk^\mu}{dP} ,
    \label{eq:pos_geo_thetao}
    \\
    \frac{d}{d\lambda} \frac{dk^\mu}{dP} = \frac{d}{dP}\left( \Gamma^\mu_{\alpha \beta} k^\alpha k^\beta\right),
    \label{eq:mom_geo_thetao}
    \\
    \frac{d}{d\lambda} \frac{d}{dP}\left(\frac{I_\nu}{\nu^3}\right) = \frac{d}{dP} \left[\left( \frac{j_{\nu}}{\nu^2}\right) - (\alpha_\nu \nu) \left(\frac{I_\nu}{\nu^3} \right) \right].
    \label{eq:rad_trans_thetao}
\end{gather}

Since the connection coefficients are functions of the photon's four-position, i.e., $\Gamma^\mu_{\alpha\beta} \equiv \Gamma^\mu_{\alpha\beta}(x^\mu)$, and considering the specific emissivity and absorption to be functions of the photon's four-position, four-momentum and the parameters, i.e., $j_\nu \equiv j_\nu(x^\mu, k^\mu, P)$, $\alpha_\nu \equiv \alpha_\nu(x^\mu, k^\mu, P)$, we can rewrite Eqs.~\eqref{eq:mom_geo_thetao} and~\eqref{eq:rad_trans_thetao} as
\begin{align}
    \frac{d}{d\lambda} \frac{dk^\mu}{dP} 
    &= \frac{d}{dP}f_1(x^\mu, k^\mu, P) \notag \\
    &= \frac{\partial f_1}{\partial x^\mu} \frac{dx^\mu}{dP}
     + \frac{\partial f_1}{\partial k^\mu} \frac{d k^\mu}{d P} + \frac{d f_1}{d P}, 
    \label{eq:mom_geo_simp_thetao} \\
    \frac{d}{d\lambda} \frac{d}{dP}\left(\hat{I}_\nu\right) 
    &= \frac{d}{dP}f_2(x^\mu, k^\mu, \hat{I}_\nu, P) \notag \\
    &= \frac{\partial f_2}{\partial x^\mu} \frac{dx^\mu}{dP}
     + \frac{\partial f_2}{\partial k^\mu} \frac{d k^\mu}{dP}
     + \frac{\partial f_2}{\partial \hat{I}_\nu} \frac{d\hat{I}_\nu}{d P} + \frac{df_2}{dP},
    \label{eq:rad_trans_simp_thetao}
\end{align}
where we have defined the invariant intensity by $\hat{I}_\nu = I_\nu/\nu^3$, and the functions $f_1$ and $f_2$ as the right hand side of Eqs.~\eqref{eq:mom_geo} and~\eqref{eq:rad_trans_inv}, respectively. We compute the partial derivatives of these functions using AD with the \textit{ForwardDiff.jl} package~\citep{RevelsLubinPapamarkou2016}. To summarize, we integrate these 4-position and 4-momentum differentials ($dx^\mu/dP$ and $dk^\mu/dP$) backwards alongside the geodesic integration, while $d\hat{I}_\nu/dP$ is integrated forward alongside the radiative transfer integration. 

\section{Numerical Methods}
\label{sec:numerical_methods}
In this section, we describe the numerical methods used to solve the equations described in Section~\ref{sec:governing_equations} and the key important technicalities of implementing \texttt{Jipole}. 

\subsection{Geodesics Integration}

We utilize the second-order Runge--Kutta (RK2) method to integrate the geodesic equation, while $dx^\mu/dP$, $dk^\mu/dP$ are integrated using the Euler method as
\begin{gather}
   \left( \frac{dx^\mu}{dP}\right)_{n + 1} = \left(\frac{dx^\mu}{dP}\right)_{n} - d\lambda\ \left(\frac{dk^\mu}{dP}\right)_{n},
   \\
    \left( \frac{dk^\mu}{dP}\right)_{n+1} = \left(\frac{dk^\mu}{dP}\right)_{n} - d\lambda\ (A_1)_{n},
\end{gather}
where $A_1$ is the right-hand side of Eq.~\eqref{eq:mom_geo_simp_thetao}. The minus signs in these expressions arise because we are integrating the geodesics backwards from the camera. 

\subsection{Intensity Integration}
Once the geodesic path is computed, we integrate Eq.~\ref{eq:rad_trans_inv} by considering that along each integration step, the averaged emissivity and absorption can be described as the average between the previous and the current step, i.e., $j_\nu^{\rm avg} = (j_\nu^{n} + j_\nu^{n+1})/2$ and $\alpha_\nu^{\rm avg} =  (\alpha_\nu^{n} + \alpha_\nu^{n+1})/2$, where $n$ is the integration step. If one assumes constant emissivity and absorption coefficients over the course of a single step, the radiative transfer equation (Eq.~\ref{eq:rad_trans_inv}) has the following analytical solution:
\begin{equation}
    \hat{I}_{n + 1} = \hat{I}_{n} e^{d\lambda \ \alpha^{\rm avg}_\nu} + \frac{j_\nu^{\rm avg}}{\alpha_\nu^{\rm avg}} (1 - e^{d\lambda \ \alpha^{\rm avg}_\nu}).
    \label{eq:Int_solve_approx}
\end{equation}
Alongside the intensity integration, we integrate $d\hat{I}_\nu/d\theta_o$ using the Euler method as
\begin{equation}
        \left( \frac{d\hat{I}}{dP}\right)_{n+1} = \left(\frac{d\hat{I}}{dP}\right)_{n} + dl_{\rm cgs}\ (A_2)_{n},
\end{equation}
where $A_2$ denotes the right-hand side of Eq.~\ref{eq:rad_trans_simp_thetao}. The step size in ${\rm cgs}$ units is $d\lambda_{\rm cgs} = d\lambda \times [L_{\rm unit} h/(m_e c^2)]$, where $L_{\rm unit}$ is the length scale dictated by the mass of the black hole, $h$ is Planck's constant in $\rm {ergs} \cdot \rm{s}$, $m_e$ is the electron mass in grams, and $c$ is the speed of light in $\rm{cm}/\rm{s}$. This scaling is performed when passing the step size to the intensity integration function, and it is necessary, as the radiative transfer equation is expressed in physical units.

\subsection{The Conjugate Gradient Method}
\label{sec:CG}

To illustrate the potential improvements that differentiable imaging enables for traditional model-fitting procedures, we present a demonstration using conjugate gradient optimization. This example is intended purely as a proof-of-concept and does not constitute a rigorous model-fitting procedure suitable for actual data analysis, as it lacks safeguards against local minima, model mis-specification, and other systematic errors.

The outcomes of the integration described in the previous section are the pixel-wise differentials of the intensity $I_\nu$, with respect to the model parameters, e.g., the observer angle $\theta_o$ or the dimensionless spin parameter $a$, paramount information for data analysis. As a simple data analysis example, in this work, to infer the best-fit parameters that reproduce a simulated image, we minimize the following cost function given by the mean squared error (MSE) between the simulated image $I_{\rm ref}$ and the model-generated image $I_{\rm test}$:
\begin{equation}
    \mathcal{C}(I_{\rm ref}, I_{\rm test}) = \frac{1}{n_xn_y} \sum_{i,j=1}^{n_x, n_y} \left(I_{\rm ref}^{(i,j)} - I_{\rm test}^{(i,j)}\right)^2,
    \label{eq:cost_function}
\end{equation}
where $n_x$ and $n_y$ are the image dimensions, and $i$ and $j$ denote pixel coordinates. 

To minimize the cost function $\mathcal{C}$, we use a modified version of the conjugate gradient (CG) method \citep{Hestenes_1952}. We first compute the gradient of the cost function with respect to the parameters, by simply using the chain rule to get:
\begin{equation}
    \frac{d\mathcal{C}}{dP} = -\frac{2}{n_xn_y} \sum_{i,j=1}^{n_x, n_y} \left(I_{\rm ref}^{(i,j)} - I_{\rm test}^{(i,j)}\right) \frac{dI^{(i,j)}_{\rm test}}{dP},
    \label{eq:grad_cost_func}
\end{equation}
where ${dI^{(i,j)}_{\rm test}}/{dP}$ is the intensity differentials obtained from the integration scheme for each pixel. At each iteration, the CG method chooses a descent direction based on both the current gradient and the previous search direction using this criterion
\begin{equation}
    \mathbf{p}_k = -\nabla \mathcal{C}_k + \beta_k \mathbf{p}_{k-1},
\end{equation}
where $\beta_k$ is the Polak--Ribi\`ere formula \citep{Polak_1969}
\begin{equation}
    \beta_k = \max\left(0, \frac{(\nabla \mathcal{C}_k - \nabla \mathcal{C}_{k-1})^\top \nabla \mathcal{C}_k}{\|\nabla \mathcal{C}_{k-1}\|^2} \right).
\end{equation}

We use a backtracking line search that satisfies the Armijo condition \citep{Armijo_1966}
\begin{equation}
    \mathcal{C}(\mathbf{x}_k + \alpha \mathbf{p}_k) \leq \mathcal{C}(\mathbf{x}_k) +\delta \alpha \nabla \mathcal{C}_k^\top \mathbf{p}_k,
\end{equation}
where $0 < \delta  < 1$ is a small real constant. If the step size $\alpha$ does not satisfy this condition, it is reduced iteratively by a factor $\beta = 0.5$, until the inequality is met or a maximum number of reductions is reached (which we have set at $10$). We have also enforced physical bounds on the parameters by projecting any out-of-bounds trial step back into the allowed domain. We stop the method when the MSE between the proposal and the targeted images falls below a targeted tolerance level.

\section{Validation of \texttt{Jipole}}
\label{sec:validation}

In this section, we validate \code{Jipole} against the \code{ipole} code~\citep{Monika_2018}\footnote{For this comparison, we have utilized the analytical expressions for the connection coefficients in \code{ipole}, rather than using the finite differences method.}, using a thin disk and five different analytical models to simulate plasma properties, in the spirit of the code comparison performed in~\cite{Gold_2020}. We have only compared the total flux density and individual simulated images, leaving the study of polarization capabilities for future work. For all these tests we consider a Kerr black hole (BH) with dimensionless spin $a$ in Boyer--Lindquist coordinates. This validation is intended solely to verify the correctness of the Julia-based implementation, and, given that \code{Jipole} and \code{ipole} employ identical algorithms, the results are expected to match exactly—up to numerical round-off—across all tests. 

\subsection{Thin Disk Comparison}
\label{sec:thin_disk_comparison}

This test model, described in~\cite{Prather_2023}, assumes that the plasma distribution is represented as a geometrically thin and optically thick accretion disk aligned with the midplane. We assume the disk orbits around a rotating black hole characterized by a dimensionless spin of $a = J/M = 0.99$, and we neglect light scattering and re-emission: we always set the intensity $I_\nu$ at the first midplane crossing of the geodesic. Under these conditions, the specific intensity $I_\nu$ follows a hardened blackbody emission:
\begin{gather}
    I_\nu = \frac{1}{n^4} B_\nu (n T_{\rm eff}),
    \\
    T_{\rm eff} = \left(\frac{F}{\sigma}\right)^{1/4},
\end{gather}
where $T_{\rm eff}$ is the effective temperature of the disk at a certain radius, $\sigma$ is the Stefan--Boltzmann constant, $F$ is the total flux, computed following \cite{Page_1974}, $n$ is the hardening factor, which we set to $n= 1.8$, and $B_\nu(T)$ is the blackbody function of temperature.

The emitted intensity follows the angular distribution for scattering from a semi-infinite atmosphere as presented in Table 24 of \cite{Chandrasekhar_1960}. The disk emission is assumed to be generated only between $r_{\rm ISCO}$ and $100\ r_g$, where $r_g = GM_{\rm BH}/c^2$, where $G$ is the gravitational constant and $M_{\rm BH}$ is the mass of the black hole. 

We assume that the gas particles in the disk move with a Keplerian angular velocity $u^\phi/u^t = 1/(r^{3/2} + a)$ and these values for the rest of the parameters of the model:
\begin{gather}
    M_{\rm BH} = 10\ M_\odot,
    \\
    D_{\rm source} = 0.05\ \rm{pc},
    \\
    \dot{M} = 0.1\ \dot{M}_{\rm edd},
    \\
    h\nu = 1 \ \rm{keV},
\end{gather}
where $D_{\rm source}$ is the source's distance to the observer, and $\dot{M}$ is the accretion rate defined in terms of the Eddington unit $\dot{M}_{\rm edd}$.

\begin{figure*}
    \centering
    \includegraphics[width=\linewidth]{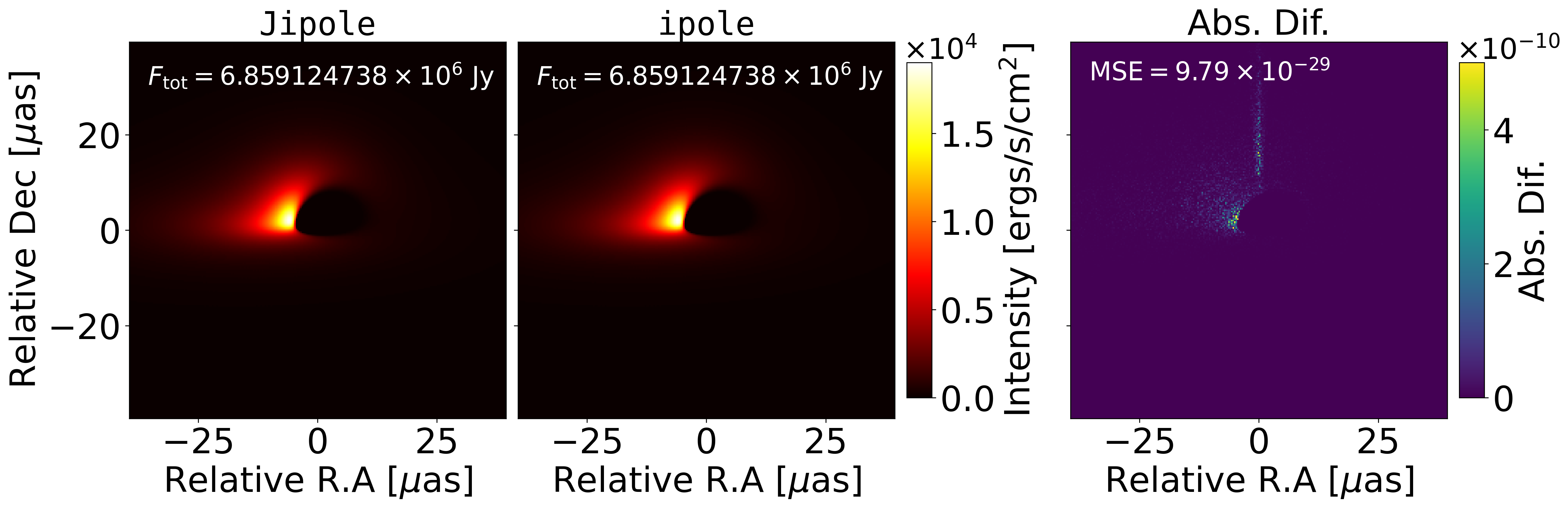}
    \caption{Image intensity map generated by \code{Jipole} (left column) and \code{ipole} (middle column) for the thin disk model described in Section~\ref{sec:thin_disk_comparison}. The right column illustrates the absolute difference between the two images. Total flux density values (see Eq.~\ref{eq:flux_density}) are overlaid as text on the intensity maps in the left and middle columns, and the NMSE (Eq.~\ref{eq:NMSE}) is overlaid on the absolute difference map in the right column.}
\label{fig:thindisk_comparison}
\end{figure*}

The resulting images of this thin disk model from both \code{Jipole} and \code{ipole} are shown in Fig.~\ref{fig:thindisk_comparison}. All the images in the validation section are of size $256\times256$ pixels. The flux density values are overlaid on each image and are calculated as
\begin{equation}
    F_{\rm tot} = \nu^3 S\sum_{(i, j)}^{(n_x, n_y)} \hat{I}_{(i,j)},
    \label{eq:flux_density}
\end{equation}
where $\nu$ is the frequency and $S$ is a scale parameter defined by $S = dx/n_x \times dy/n_y \times 1/D_{\rm source}^2$.

As shown in the right panel of Fig.~\ref{fig:thindisk_comparison}, which displays the absolute difference of the computed images, both intensity maps generated by \code{Jipole} and \code{ipole} demonstrate a high level of agreement, with total flux density values matching up to fifteen significant digits. The absolute difference map reveals only slight discrepancies, with errors around $10^{-10}$, which are negligible compared to the typical pixel intensities ranging from $10^3$ to $10^4$. These residuals, which are several orders of magnitude below the intensity signal, are most likely arising from floating-point round-off errors. With text, in the third column of Fig.~\ref{fig:thindisk_comparison} we  show the normalized mean squared error (NMSE) computed as 
\begin{equation}
\text{Normalized MSE} = \frac{\sum \left( I_{\text{Jipole}} - I_{\text{ipole}} \right)^2}{\sum I_{\text{ipole}}^2},
\label{eq:NMSE}
\end{equation}
where $I_{\rm Jipole}$ is the intensity calculated using \code{Jipole} and $I_{\rm ipole}$ is the intensity calculated using \code{ipole}. For this test case, the MSE was $\sim 9.79 \times 10^{-29}$.

\subsection{Analytical Parameterized Matter Distributions}
\label{sec:analytic_model_comparison}

Let us now validate \texttt{Jipole} using the parameterized matter distributions models (models 1-5) showcased in Section 3.2 of \cite{Gold_2020}, where simple emissivities and absorptivities are prescribed. In these analytical models, both $j_\nu$ and $\alpha_\nu$ are given in the fluid-frame, and therefore the fluid 4-velocity has to also be modeled. We consider a camera at a distance of $1000 \ r_g$ with a field of view (FOV) spanning $30\ r_g$ in both $x$ and $y$ directions, observing a BH with mass $M_{\rm BH} = 4.063\times10^6 \ M_\odot$, located at a distance $D_{\rm source} = 7.78 \ \rm{kpc}$. For these models the number density of the fluid is given by
\begin{equation}
    n = n_0 \exp\left\{ -\frac{1}{2} \left[ \left( \frac{r}{10} \right)^2 + z^2 \right] \right\},
\end{equation}
where $z = h \cos\theta$, $n_0$ is a reference value and $h$, here, represents the scale height of the disk. The angular momentum profile is given by
\begin{equation}
    l = \left( \frac{l_0}{1 + R} \right) R^{1+q},
\end{equation}
where $l_0$ acts as a switch, being either $0$ or $1$, $R$ is the cylindrical radius, $R=r\sin(\theta)$, and $q = 0.5$ is the power-law index. The covariant 4-velocity of the fluid is given by
\begin{equation}
    u_\mu = \tilde{u}(-1,\, 0,\, 0,\, l),
\end{equation}
where the normalization factor ($\tilde{u}$) is
\begin{equation}
    \tilde{u} = \left[ -g^{tt} - 2g^{t\phi}l + g^{\phi\phi}l^2 \right]^{-1/2},
\end{equation}
such that $u^\mu u_\mu = -1$. For these models, the invariant emissivity is given by
\begin{equation}
    j_\nu = \frac{1}{\nu^2}C n \left( \frac{\nu}{\nu_p} \right)^{-\alpha},
\end{equation}
where $Cn_0 = 3 \times 10^{18} \rm{ergs\ cm^{-3}\ s^{-1}\ Hz^{-1}\ sr^{-1}}$, the power-law index $\alpha$ is a free parameter, and we set $\nu_p = 230\ \rm{GHz}$. The invariant absorptivity of the fluid, the last ingredient of these analytical models, is given by
\begin{equation}
    \alpha_\nu = \nu A C n \left( \frac{\nu}{\nu_p} \right)^{-(\beta + \alpha)},
\end{equation}
where $\beta = 2.5$ and $A$ are also free parameters.

\begin{figure*}
    \centering
    \includegraphics[width=\linewidth]{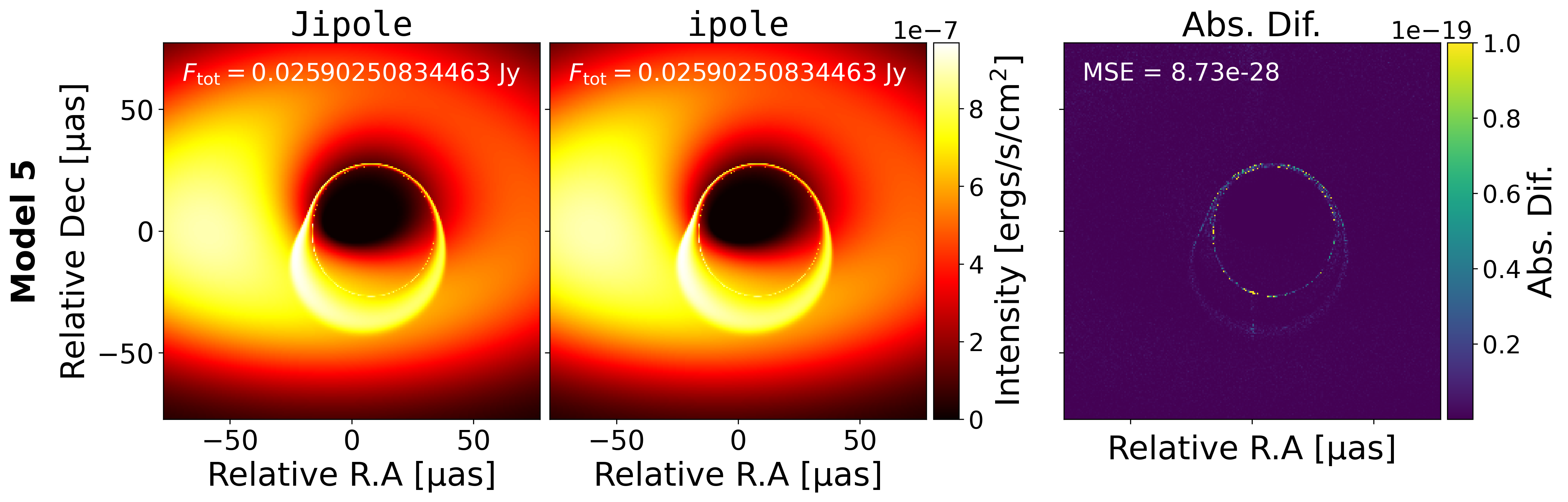}
    \caption{Intensity images generated by \code{Jipole} (left column) and \code{ipole} (middle column), alongside their absolute difference (right column), for the analytical Model 5 presented in \cite{Gold_2020}. The total flux density ($F_{\rm tot}$) for each image is annotated in the top-left corner of the \code{Jipole} and \code{ipole} panels, while the NMSE (Eq.~\ref{eq:NMSE}) on the top-left corner of the right panel.}
\label{fig:analytical_comparison}
\end{figure*}

We follow the same parameter setup as outlined in \cite{Gold_2020},  meaning the free parameters of these models are $a$, $A$, $\alpha$, $h$, and $l_0$. To illustrate the agreement between \texttt{ipole} and \texttt{Jipole}, in Fig.~\ref{fig:analytical_comparison} we present images for Model 5 (following the nomenclature used in~\cite{Gold_2020}), which we choose as an example testing emission and absorption. This model characterizes the disk with the parameters $A = 10^6$, $\alpha = 0$, $h = 100/3$, $l_0 = 1$, and $a = 0.9$.

As expected, the results from both codes agree down to machine precision errors across these analytical models. The total flux density values, computed according to equation~\ref{eq:flux_density} with $\nu = 230\ \rm{GHz}$, are annotated in the top-left corner of each panel. The absolute difference images indicate negligible error in most pixels, except near the photon orbit, where errors can reach magnitudes on the order of $\sim 10^{-17}$. The errors observed stem from truncation errors, as photons trapped near the photon ring undergo many more integration steps, resulting in cumulative numerical inaccuracies. This comparison has been conducted for all five analytical tests described in \cite{Gold_2020}, and all tests demonstrated a consistent level of agreement similar to that shown for Model 5.

\subsection{Automatic Differentiation vs Finite Differences}

With the correctness of the Julia-based implementation of \texttt{ipole} established, we can now proceed to compute the intensity differentials. Before doing so, however, we must also verify the accuracy of the derivatives obtained via AD. To validate our AD approach, we compare the computed derivatives against FD approximations. We will compute the FD derivatives using the central difference formula
\begin{equation}
    \frac{\partial I}{\partial p} = \frac{I(p + \epsilon) - I(p - \epsilon)}{2\epsilon},
\end{equation}
where $p$ represents the parameter of interest, either the spin parameter $a$ or the observer inclination angle $\theta_o$, and $\epsilon$ is a small perturbation to the chosen parameter, here chosen to be $\epsilon = 10^{-5}$.

\begin{figure*}[h!]
    \centering
    \includegraphics[width=\linewidth]{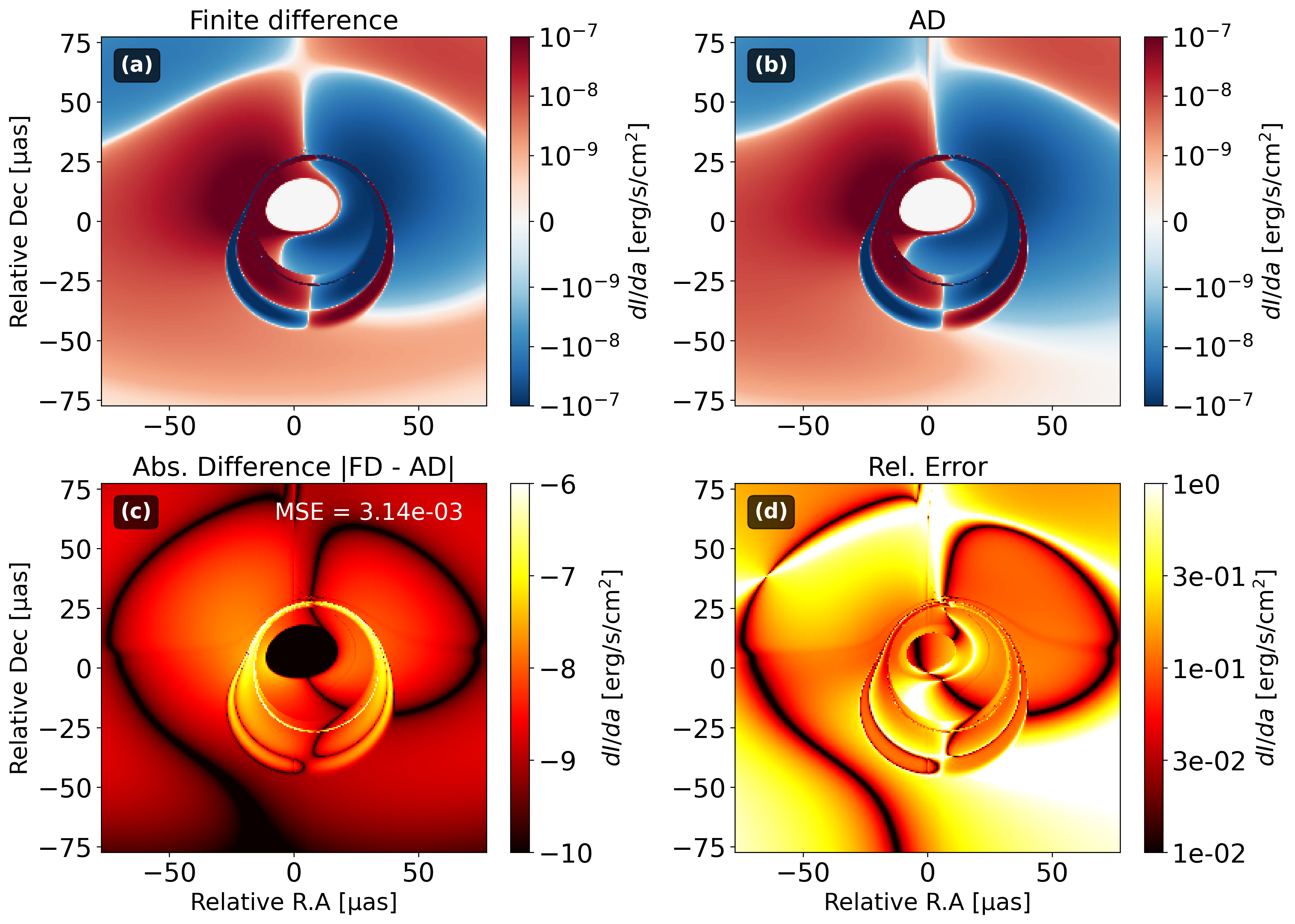}
    \caption{Image comparison of the intensity differential $dI/da$ for each pixel generated by \code{Jipole}. Panel (a): $dI/da$ computed using the FD method, shown in symmetric logarithmic scale. Panel (b): $dI/da$ computed using AD, shown in symmetric logarithmic scale. Panel (c): Absolute difference between the FD and AD results, in logarithmic scale; the NMSE (Eq.~\ref{eq:NMSE}) is indicated in the top-right corner. Panel (d): Relative difference between the FD and AD results, in logarithmic scale. These intensity differentials display the characteristic lensing behavior of black holes: nested rings that are successively rotated and demagnified.}
\label{fig:spin_comparison}
\end{figure*}

In Fig.~\ref{fig:spin_comparison}, we show $dI/da$ for each pixel computed via FD and AD (panels a, b), along with their absolute and relative differences (panels c, d), using the analytical Model 5 used in the previous section. Panels (a) and (b) reveal that both methods produce very similar images. As expected, $dI/da$ grows significantly near the photon ring, where trajectories are highly sensitive to the black hole’s spin; even small changes in spin can strongly alter the brightness and structure of these pixels. This intensity differential naturally reveals the characteristic lensing behavior of Kerr black holes~\citep{Gralla_Lupsasca_2020}, exhibiting a banded structure (lensing bands) in the photon ring~\citep{CardenasAvendano_2022} that is successively demagnified and rotated in subsequent images.

In Fig.~\ref{fig:spin_comparison} (c), the absolute difference is generally about an order of magnitude smaller than the typical intensity values in each image, except near the photon ring. This behavior is expected, as already mentioned above: in the FD approach, small numerical perturbations in the parameter lead to slightly different ray trajectories, producing larger deviations precisely where the intensity varies most rapidly. In Fig.~\ref{fig:spin_comparison} (d), we also show the relative error
\begin{equation}
    \epsilon_{\rm rel} = \frac{\lvert FD - AD\rvert}{\lvert FD\rvert},
\end{equation}
where $FD$, in this case, means $dI/dP$ as computed by FD, and $AD$ means $dI/dP$ as computed by AD, for the parameter $P$. 

\begin{figure*}[h!]
    \centering
    \includegraphics[width=\linewidth]{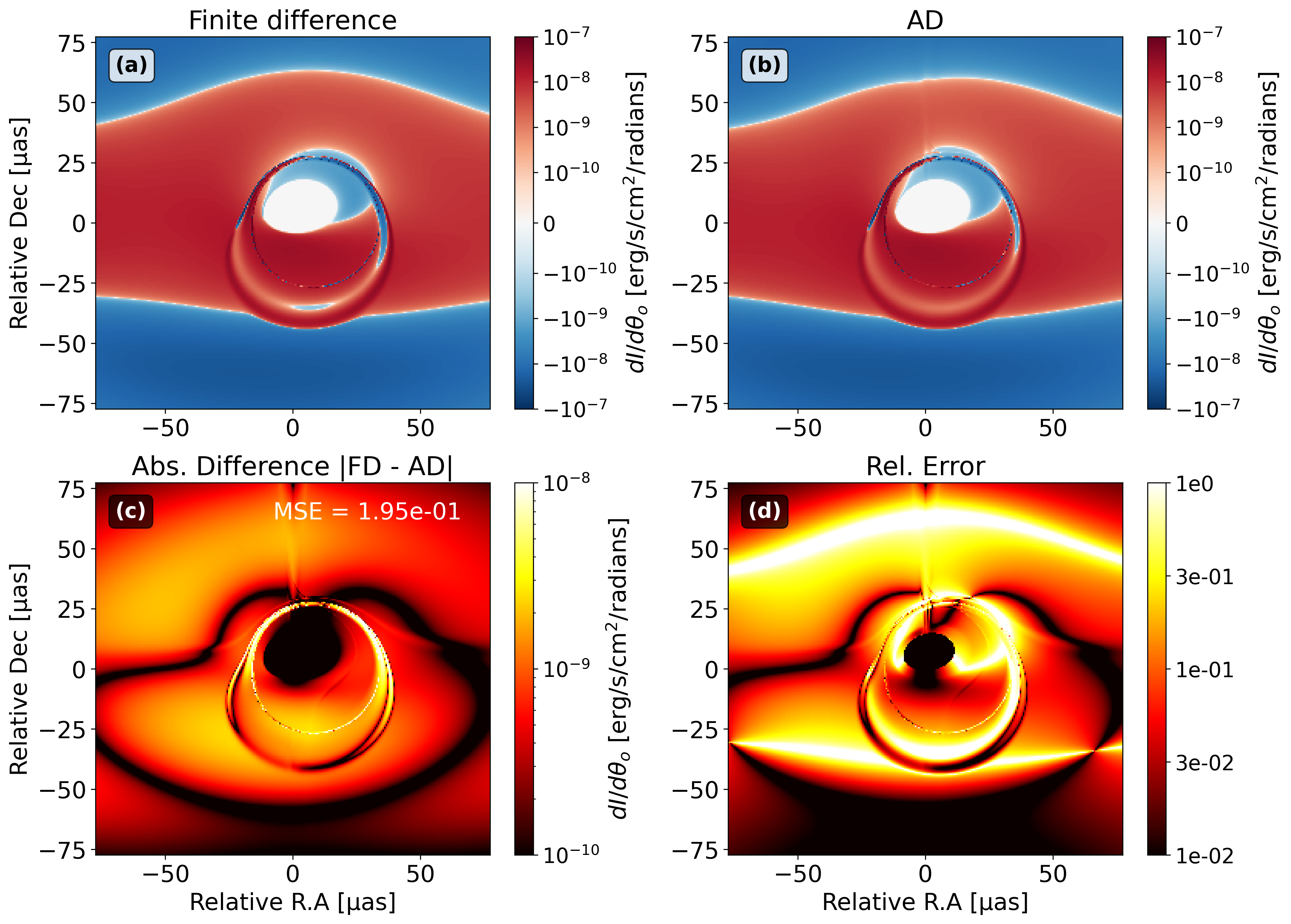}
    \caption{Image comparison of $dI/d\theta_o$ for each pixel generated by \code{Jipole}. Panel (a): $dI/d\theta_o$ computed using the FD method, shown in symmetric logarithmic scale. Panel (b): $dI/d\theta_o$ computed using AD, shown in symmetric logarithmic scale. Panel (c): Absolute difference between the FD and AD results, in logarithmic scale; the NMSE (Eq.~\ref{eq:NMSE}) is indicated in the top-right corner. Panel (d): Relative difference between the FD and AD results, in logarithmic scale.}
\label{fig:theta_comparison}
\end{figure*}

In panel (d), the error reaches its maximum in regions where $dI/da$ is very small. The white contours indicate regions with a sign inversion, as seen in panels (a) and (b), while the black contours mark areas where the AD and FD results are in close agreement, as highlighted in panel (c).
Figure~\ref{fig:theta_comparison} presents the corresponding test for $dI/d\theta_o$. In this case, both methods again produce very similar images—indeed, even more similar—owing to the smoother response of geodesic trajectories to changes in viewing angle compared to spin. As shown in Figure~\ref{fig:theta_comparison}(c), the absolute difference is, overall, much smaller than the $dI/d\theta_o$ values in panels (a) and (b). In Figure~\ref{fig:theta_comparison}(d), most of the discrepancy between the FD and AD results arises at the edges of the emission region, where the intensity drops sharply to zero, and near the photon ring.

The aforementioned discrepancies between these FD and AD results are expected. The FD method requires computing the intensity $I$ at two distinct parameter values, which, in the context of ray-tracing, entails integrating along two different geodesic paths. When the parameter $p$ is perturbed, the null geodesics connecting the observer to the emission region change, sampling different regions of the accretion flow and traversing different spacetime trajectories. In contrast, AD computes derivatives by propagating differential information along a \emph{single} geodesic path, without the need for separate geodesic integrations. Consequently, the two approaches fundamentally measure different quantities and may not yield identical results, particularly in regions where geodesics exhibit high sensitivity to initial conditions, such as near the photon ring, where small parameter changes can cause geodesics to sample dramatically different emission regions. This difference is intrinsic to the methods themselves: AD captures the derivative along a fixed trajectory, while FD approximates it through sampling neighboring trajectories. The agreement between methods in smooth regions validates the AD implementation, while localized discrepancies near caustics reflect the physical reality of geodesic sensitivity in strong gravity.

\section{Data Analysis Application}
\label{sec:results}

We now combine the intensity differentials $dI/da$ and $dI/d\theta_o$ with the gradient-based optimization method described in Section~\ref{sec:CG} to carry out a simple data analysis experiment. For these tests, we inject a simulated image—initially without blurring or noise—and attempt to recover its underlying parameters, focusing exclusively on these geometrical ones, spin and inclination, for which the differentials have been computed.

Note that in the context of real observations with systematic and stochastic errors, a Bayesian approach would be required to characterize posterior distributions and thereby gauge certainty about the fitted parameters. Here, we aim only to gauge the robustness of using the differential image to recover parameters from a simple model, to understand whether and how it reacts to blurring and noise. As we discussed in Sec.~\ref{sec:CG}, this setup is illustrative rather than intended for actual model fitting. In particular, \code{Jipole} in this usage does not include mechanisms to mitigate local minima, model mis-specification, or other systematic effects present in real observations. Ongoing work is focused on integrating \code{Jipole} into full sampling frameworks, such as \code{Comrade.jl} \citep{Tiede_2022}, to enable a more rigorous inference under realistic conditions.

To evaluate the efficiency of the fitting procedure, we begin with the analytical Model 5 described in Section~\ref{sec:analytic_model_comparison}. Using the camera parameters specified in that section, we set the target values to $\theta_o = 60^\circ$ and $a = 0.9$ and generate a $128 \times 128$ image for the fitting. We first fit $\theta_o$ alone, then $a$, and then perform a joint fit of both parameters.

\subsection{Inferring the spacetime parameters in Model 5}

\begin{figure}
    \centering
    \includegraphics[width=\linewidth]{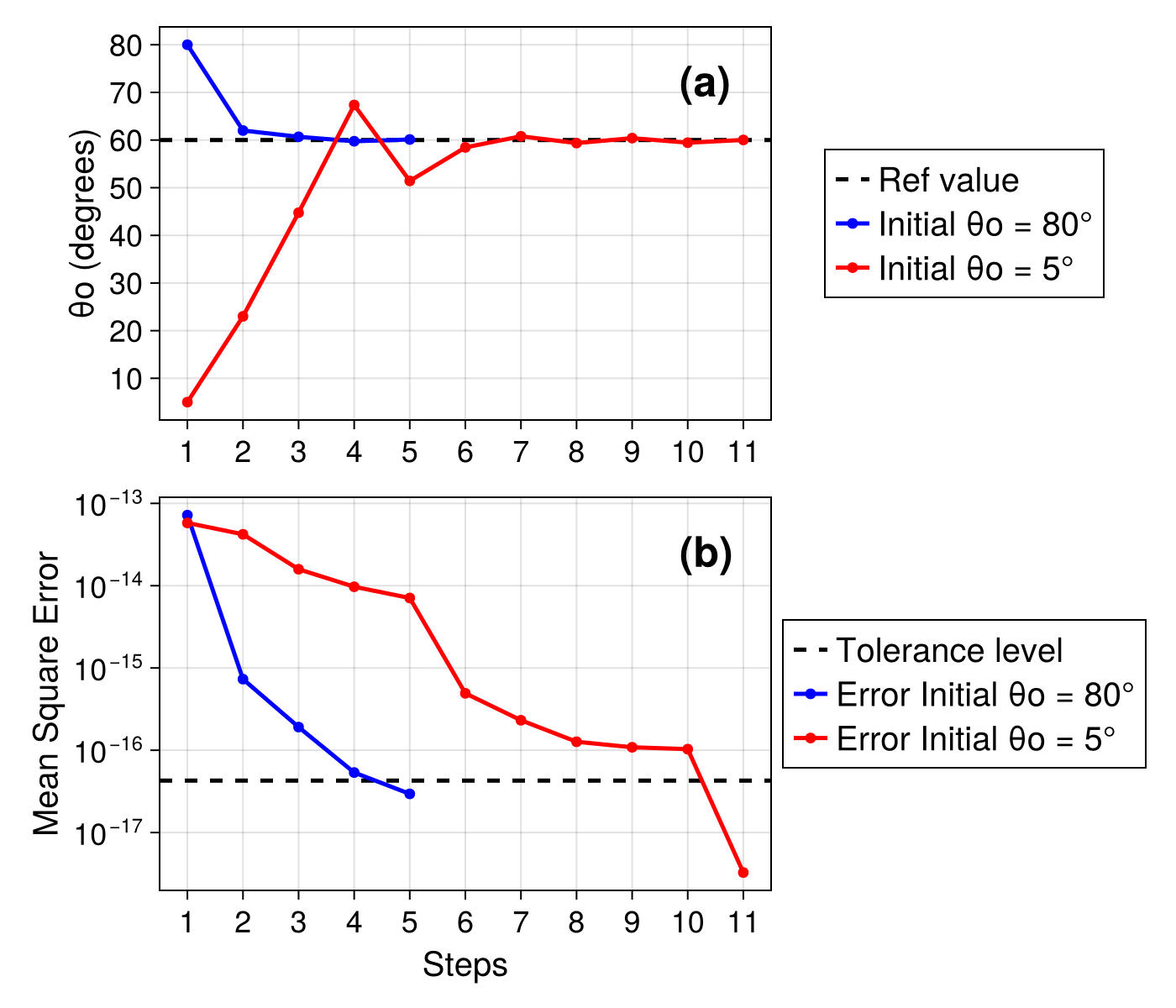}
    \caption{Convergence of the value of the observer’s inclination fitting using conjugate gradient optimization for the analytical Model 5. Panel (a): Evolution of the inclination angle $\theta_o$ during iterative fitting, starting from two initial conditions—$\theta_o = 80^\circ$ (blue) and $\theta_o = 5^\circ$ (red)—and converging toward the reference value of $60^\circ$ (dashed black line). Panel (b): The MSE during the $\theta_o$ fitting, showing convergence to the tolerance threshold of $4 \times 10^{-17}$ (dashed black line).}
\label{fig:model5_thfitting}
\end{figure}

For the observer angle, we consider two initializations: one above the target value at $\theta_o = 80^\circ$, and one below at $\theta_o = 5^\circ$. Figure~\ref{fig:model5_thfitting}(a) shows the evolution of $\theta_o$ during the CG iterations. The run initialized at $\theta_o = 80^\circ$ (blue line) converges to $\theta_o = 60.117^\circ$ after $5$ steps, while the run starting from $\theta_o = 5^\circ$ (red line) converges to $\theta_o = 60.022^\circ$ after $11$ steps. Panel (b) presents the MSE for both runs over the course of the iterations. The stopping tolerance was set to $4\times 10^{-17}$.

\begin{figure}
    \centering
    \includegraphics[width=\linewidth]{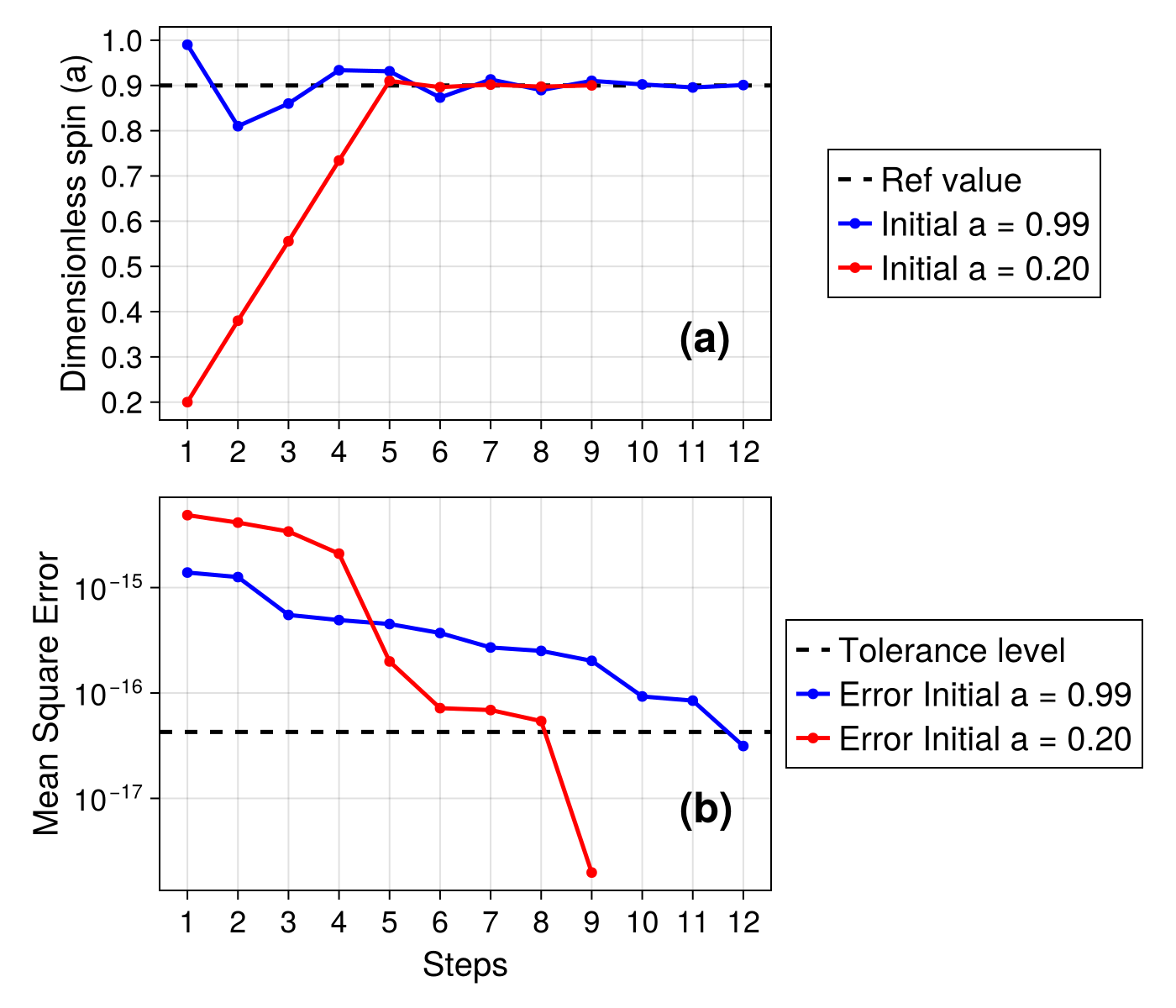}
    \caption{Convergence of the spin parameter fitting using conjugate gradient optimization for the analytical Model 5. Panel (a): Evolution of the dimensionless spin $a$ during iterative fitting, starting from two initial conditions—$a = 0.99$ (blue) and $a = 0.20$ (red)—and converging toward the reference value $a = 0.9$ (dashed black line). Panel (b): The MSE during the $a$ fitting, showing convergence to the tolerance threshold of $4 \times 10^{-17}$ (dashed black line).}
\label{fig:model5_afitting}
\end{figure}

For the dimensionless spin parameter, we also consider two initializations: one above the expected target at $a = 0.99$, and one below at $a = 0.2$. Figure~\ref{fig:model5_afitting}(a) illustrates the evolution of $a$ during the CG iterations. The run initialized at $a = 0.99$ (blue line) converges to $a = 0.9009$ after $12$ steps, while the run starting from $a = 0.2$ (red line) reaches $a = 0.9002$ after $9$ steps. Panel (b) shows the corresponding MSE for both runs. The convergence criterion was again set to $\sim 4 \times 10^{-17}$.

\begin{figure}
    \centering
    \includegraphics[width=\linewidth]{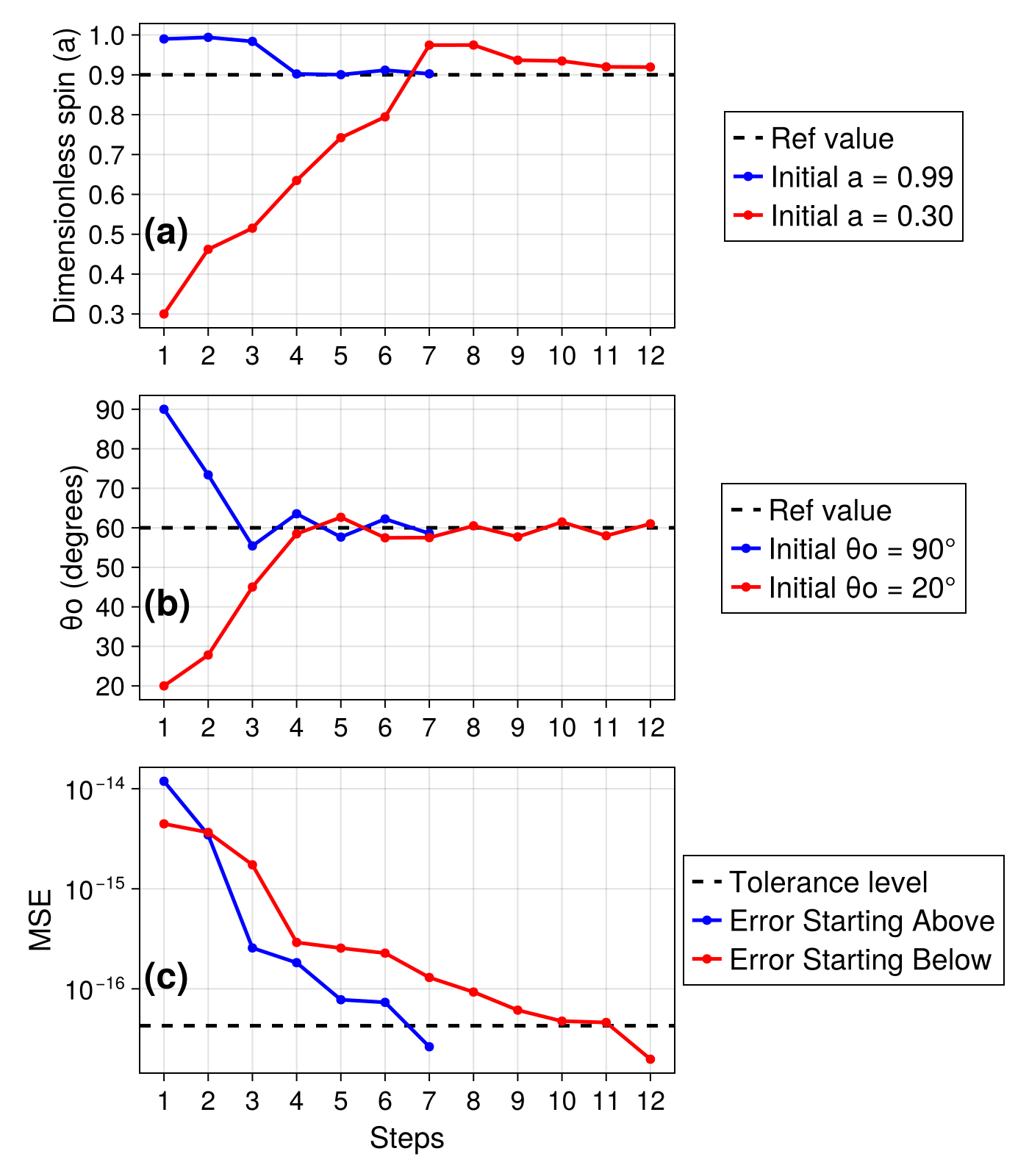}
    \caption{Convergence of the joint fitting of spin and inclination using conjugate gradient optimization for the analytical Model 5. Panel (a): Evolution of the dimensionless spin $a$ during iterative fitting, starting from two initial conditions—$a = 0.99$ (blue) and $a = 0.30$ (red)—and converging toward the reference value $a = 0.9$ (dashed black line). Panel (b): Evolution of the inclination angle $\theta_o$ during iterative fitting, starting from two initial conditions—$\theta_o = 90^\circ$ (blue) and $\theta_o = 20^\circ$ (red)—and converging toward the reference value $\theta_o = 60^\circ$ (dashed black line). Panel (c): The MSE during the joint fitting, showing convergence to the tolerance threshold of $4 \times 10^{-17}$ (dashed black line).}
\label{fig:model5_bothfitting}
\end{figure}

Lastly, we perform a simultaneous fit of both parameters, starting from $(a, \theta_o) = (0.3, 20^\circ)$ and $(a, \theta_o) = (0.99, 90^\circ)$. The results are shown in Figure~\ref{fig:model5_bothfitting}, where panels (a) and (b) depict the evolution of $a$ and $\theta_o$, respectively, and panel (c) presents the MSE throughout the iterations. In all these cases, the algorithm converges to the correct values within a few steps.

\subsection{Inferring the spacetime parameters in Model 5 with Added Blurring}

Let us now introduce an additional layer of complexity—moving toward a more realistic scenario—by simulating the finite resolution of observational data. Specifically, we apply an imaging blur to the ground-truth image using the \textit{ImageFiltering.jl} package with a Gaussian kernel, corresponding to a $20\ \rm{\mu as}$ resolution. We now repeat the parameter-fitting analysis using this blurred image as the target.

\begin{figure}
    \centering
    \includegraphics[width=\linewidth]{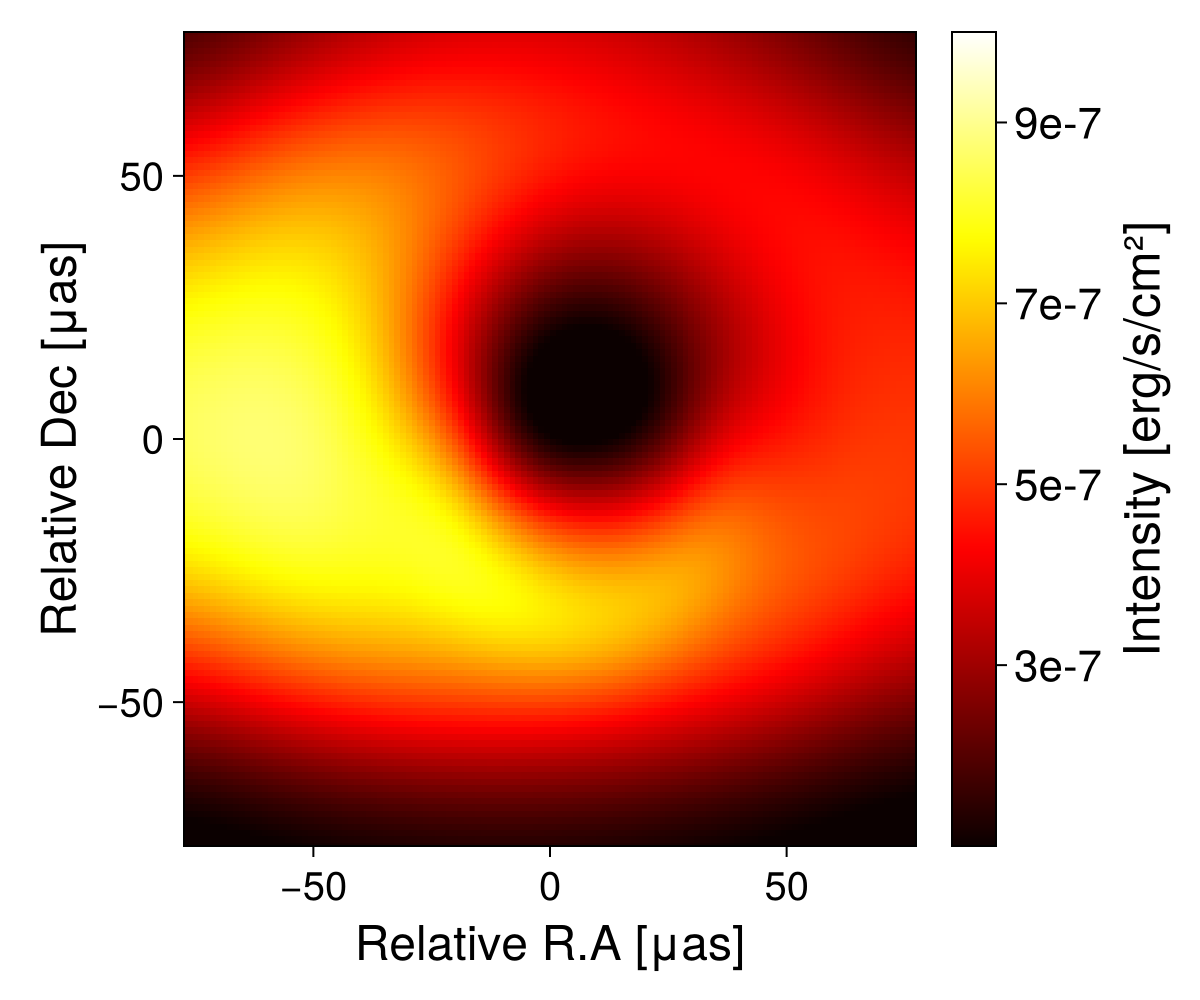}
    \caption{Blurred image intensity colormap for the analytical Model 5 described in Section~\ref{sec:analytic_model_comparison} with parameters $\theta_o = 60^\circ$ and $a = 0.9$.}
\label{fig:blurred_image}
\end{figure}

\begin{figure}
    \centering
    \includegraphics[width=\linewidth]{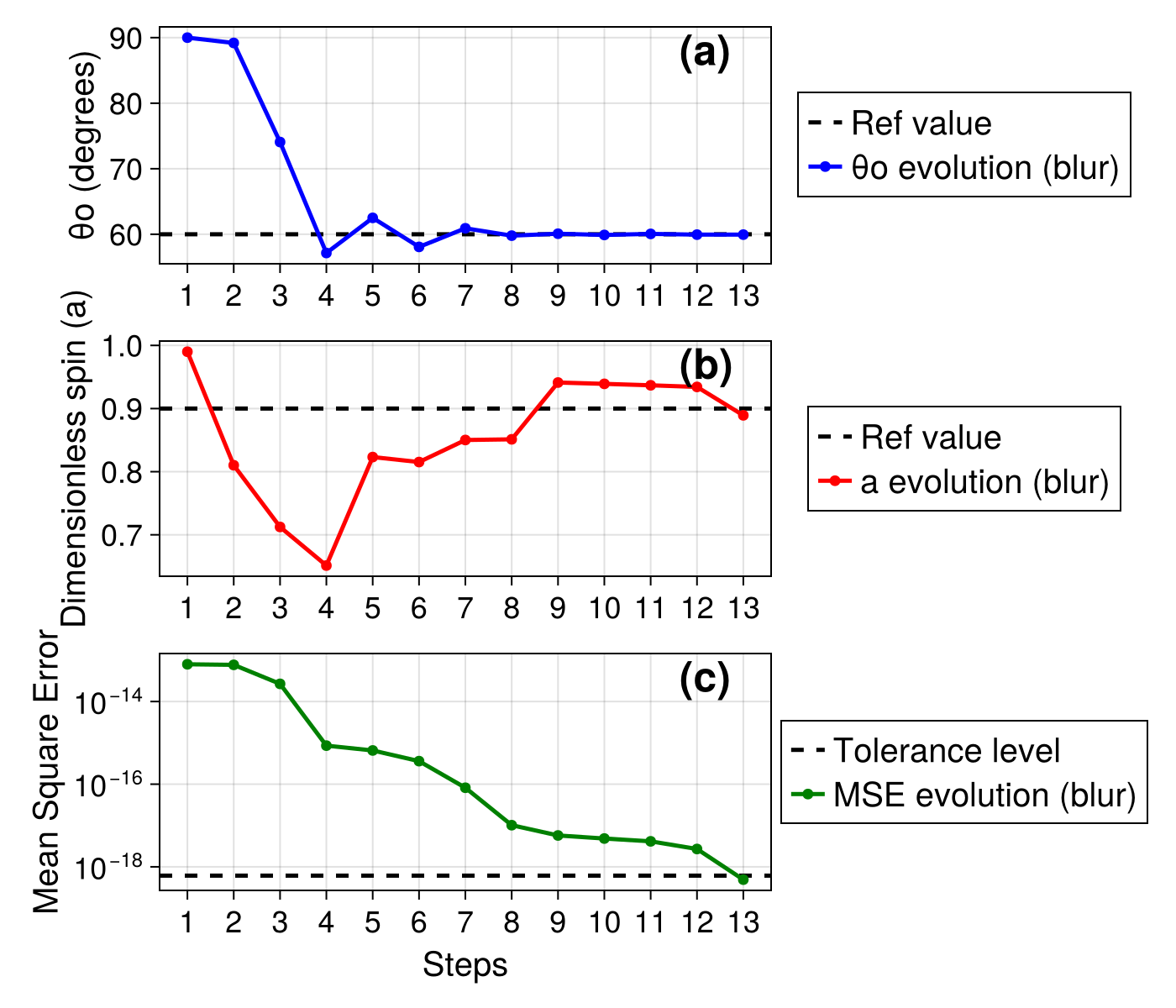}
    \caption{Convergence analysis of both parameters using conjugate gradient optimization for the blurred analytical model. Panel (a) shows the convergence of the observer's inclination $\theta_o$. The dashed black line represents the targeted value of $\theta_o = 60^\circ$. Panel (b) depicts the convergence of the dimensionless spin parameter $a$.  The dashed black line represents the targeted value of $a = 0.9$. Panel (c) showcases the mean square error for the fittings.  The dashed black line represents the tolerance level to stop the iteration at $\sim 6 \times 10^{-19}$}
\label{fig:blurred_fitting}
\end{figure}

The procedure is as follows: we generate an image for $\theta_o = 60^\circ$ and $a = 0.9$ based on Model 5, which we will treat as the ground-truth or targeted image. This image is then convolved with a Gaussian kernel $G(\sigma)$ where $\sigma$ represents the standard deviation parameter to simulate instrumental resolution 
\begin{gather}
    I_{\rm{blur}} = I_{\rm{truth}} * G({\sigma}), \\
    \sigma = \frac{\frac{N}{2\theta_{\mu\rm{as}}} \cdot 20}{2\sqrt{2\ln 2}}, 
\label{eq:blurring}
\end{gather}
where $*$ denotes convolution, $\theta_{\mu\rm{as}}$ is the field of view in microarcseconds, and $N$ is the number of pixels along one dimension of the image array. The resulting blurred target image, shown in Figure~\ref{fig:blurred_image}, is then used as input to the conjugate gradient optimization algorithm. At each iteration, the newly generated trial image and its derivatives are blurred using the same procedure. The optimization proceeds until the MSE falls below $\sim 6 \times 10^{-19}$. Figures~\ref{fig:blurred_fitting}(a-c) show the evolution of $\theta_o$, $a$, and the MSE, respectively. As in the previous section, the initial parameters are set to $(a, \theta_o) = (0.99, 90^\circ)$. The observer inclination converged to $\theta_o = 59.945^\circ$, while the black hole spin to $a = 0.889$, i.e., a successful recovery of these parameters in only a few steps (i.e., image computations).

\subsection{Inferring the spacetime parameters in Model 5 with Added Blurring and Noise}
\label{sec:noise_fitting}
We now take a further step in complexity by adding noise to the injected image. In this case, the fitting procedure follows the same approach described in the previous sections, but now it is applied to data generated from the analytical model after introducing both blurring and noise. The image is first convolved with a Gaussian kernel $G_{\sigma}$, as described in Eq.~\eqref{eq:blurring}, and then transformed to the Fourier (visibility) domain, $V = \mathcal{F}{ I_{\rm{blur}} }$, where $\mathcal{F}$ denotes the two-dimensional Fourier transform. Adding noise in the Fourier domain is preferable for interferometric simulations, as the measured quantities in VLBI are the complex visibilities rather than the image pixels themselves.

To quantify the noise level, we first compute a characteristic amplitude scale for the visibilities using the root-mean-square (RMS) value,
\begin{equation}
    \rm{rms}_{\rm{vis}} = \rm{std}\left( |V| \right).
\end{equation}
where $\mathrm{std}(\cdot)$ denotes the standard deviation. For the analysis in this section, we adopt a signal-to-noise ratio (SNR) of $\mathrm{SNR} = 15$, from which the noise standard deviation is
\begin{equation}
    \sigma_{\rm{noise}} = \frac{\rm{rms}_{\rm{vis}}}{\rm{SNR}} \, .
\end{equation}
Complex Gaussian noise is then generated as $N = \sigma_{\rm{noise}} \left[ \mathcal{N}(0,1) + i \, \mathcal{N}(0,1) \right]$, where $\mathcal{N}(\mu, \sigma)$ denotes a real-valued Gaussian random distribution with mean $\mu$ and standard deviation $\sigma$. The noise is then added to the visibility as $V_{\rm{noisy}} = V + N$.

Lastly, the noisy visibility is transformed back to the image domain as $I_{\mathrm{noisy}} = \mathrm{Re} \left\{ \mathcal{F}^{-1} \left[ V_{\mathrm{noisy}} \right] \right\}$, where $\mathcal{F}^{-1}$ denotes the inverse Fourier transform and $\mathrm{Re}(\cdot)$ extracts the real component. The resulting $I_{\mathrm{noisy}}$ serves as the input to the fitting procedure, in which the conjugate gradient method is used to recover the model parameters.

\begin{figure}
    \centering
    \includegraphics[width=\linewidth]{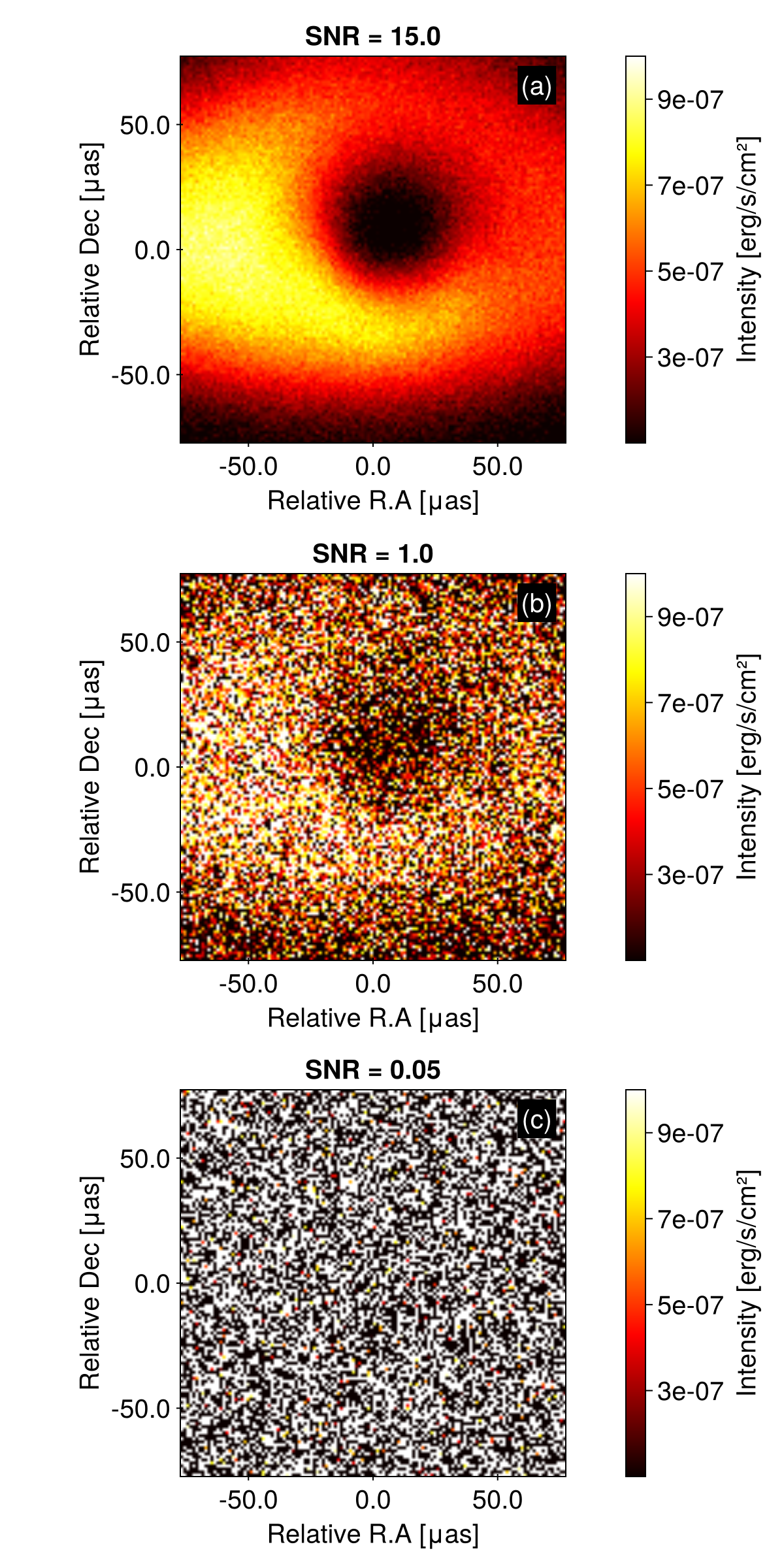}
    \caption{Resulting noisy images for the blurred analytical model described in Section~\ref{sec:analytic_model_comparison}. The blurring is applied according to Eqs.~\ref{eq:blurring}, and noise is added following the procedure in Section~\ref{sec:noise_fitting}. Panels (a)-(c) show the resulting images for signal-to-noise ratios of $15$, $1$, and $0.05$, respectively.}
\label{fig:noisy_blurred_colormaps}
\end{figure}

For this test, we consider three different signal-to-noise ratios: $\mathrm{SNR} = (15, 1, 0.05)$. The corresponding intensity maps are shown in Figure~\ref{fig:noisy_blurred_colormaps}(a-c). The noise levels for $\mathrm{SNR} = 1$ and $\mathrm{SNR} = 0.05$ are unrealistically high and are included only to test the robustness of the fitting method. In panel (b) ($\mathrm{SNR} = 1$), some underlying disk structure is still discernible, whereas in panel (c) ($\mathrm{SNR} = 0.05$) this structure is completely lost.

\begin{figure}
    \centering
    \includegraphics[width=\linewidth]{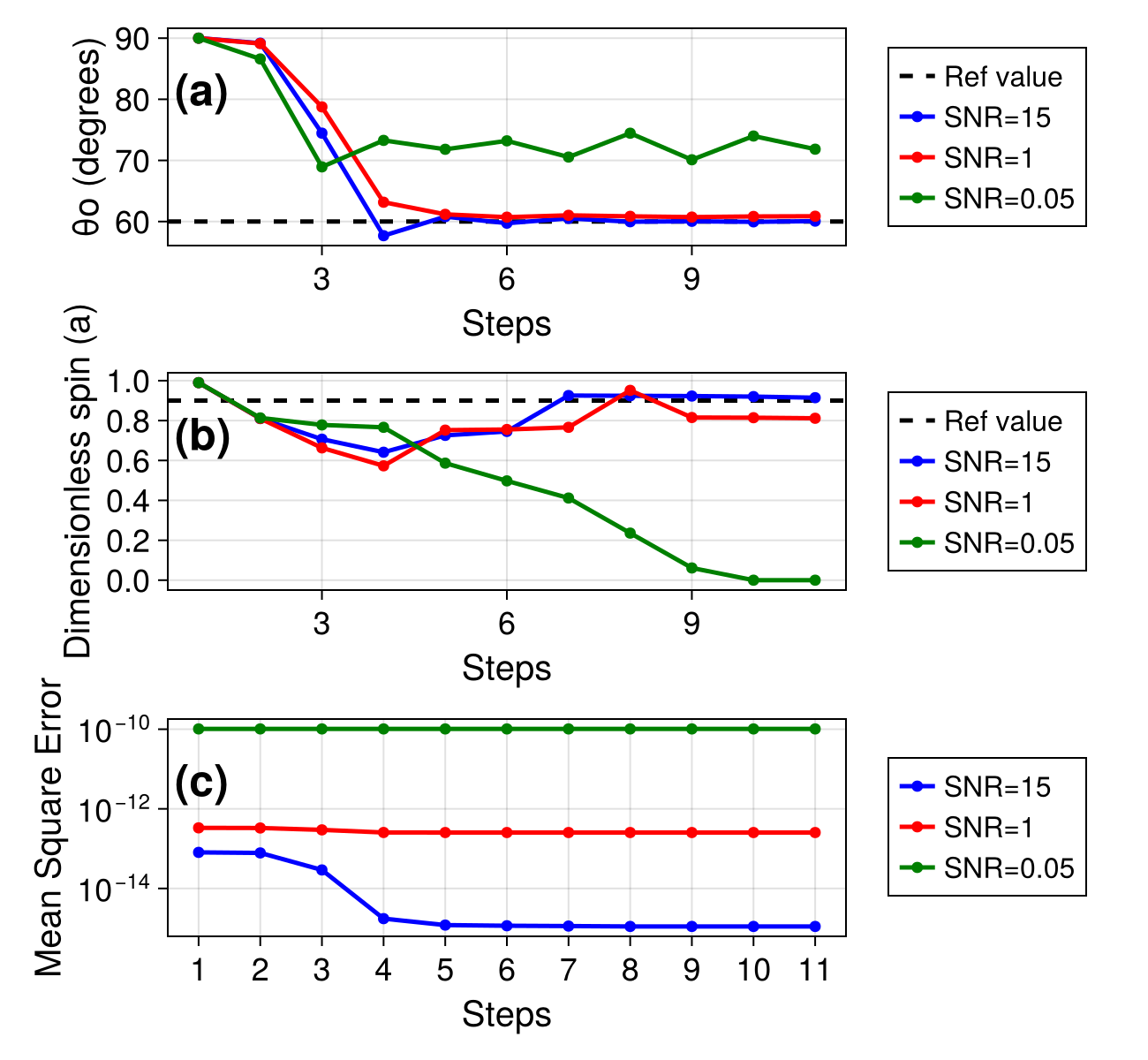}
    \caption{Evolution of the conjugate gradient fitting for blurred, noisy images with three different signal-to-noise ratios: $15$ (blue), $1$ (red), and $0.05$ (green). The black dashed lines indicate the reference values for $\theta_o$ and $a$. Panel (a): Evolution of the observer inclination over the iterations. Panel (b): Evolution of the dimensionless spin over the iterations. Panel (c): Evolution of the MSE over the iterations.}
\label{fig:noisy_fitting}
\end{figure}

The fitting evolution for the noisy images is shown in Figure~\ref{fig:noisy_fitting}. All three cases start from the same initial parameters, $(\theta_o, a) = (90^\circ, 0.99)$. For the high-SNR case ($\mathrm{SNR} = 15$), the fit correctly converges to $\theta_o \approx 60.07^\circ$ and $a \approx 0.914$. In the moderate-noise case ($\mathrm{SNR} = 1$), the recovered values are $\theta_o \approx 60.87^\circ$ and $a \approx 0.811$. For the lowest-SNR case ($\mathrm{SNR} = 0.05$), the fit yields $\theta_o \approx 71.84^\circ$ and $a = 0.0$, as expected for such a low SNR case. As shown in Figure~\ref{fig:noisy_fitting}(c), the final converged error increases consistently as the signal-to-noise ratio decreases.

Uncertainty in the estimated parameters for all the fittings performed above can be assessed by, for example, propagating the covariance structure derived from the gradients, employing resampling-based techniques, or adopting Bayesian approaches that provide posterior distributions rather than point estimates. We leave such uncertainty quantification for future work.

\section{Conclusions}
\label{Sec:Conclusions}

In this paper, we present \code{Jipole}, a differentiable code written in Julia for radiative transfer in curved spacetimes. As a proof of concept, we have only focused on computing the derivatives of the observed intensity with respect to the observer’s inclination angle and the black hole spin using automatic differentiation. This methodology can be easily expanded to include other parameters of the models.

We validated \code{Jipole} by comparing its results against those from \code{ipole} for both a thin-disk model and the simple analytical model presented in \cite{Gold_2020}, obtaining an \emph{exact}, up to round-off error, match in all cases. We further compared the intensity derivatives computed via automatic differentiation with those obtained using finite-difference methods. As expected from the different formulations for these two types of derivatives, the two approaches do not agree to round-off precision. However, they exhibit close agreement: the differences are small in magnitude, with $\mathrm{MSE} \sim 2 \times 10^{-1}$ for $dI/d\theta_o$ and $\mathrm{MSE} \sim 3 \times 10^{-3}$ for $dI/da$, and are concentrated near sharp intensity gradients, as shown in Figures~\ref{fig:spin_comparison} and~\ref{fig:theta_comparison}.

Furthermore, we demonstrated parameter fitting on the analytical plasma model using a conjugate gradient method. Fits were performed for three different datasets: the original (“ground truth”) model, a blurred version of the ground truth, and a noisy, blurred version of the ground truth, thereby simulating increasingly realistic observational conditions. The fitting procedure converged within a few number of iterations ($\sim 10$) for all datasets, with the exception of the case where the signal-to-noise ratio was $0.05$. In this extreme regime, where the noise amplitude is twenty times larger than the signal, convergence was not achieved, as expected.

These tests point to the applicability of \code{Jipole} to parameter estimation problems under a range of data quality scenarios. Though, applying \code{Jipole} to analyze real VLBI data will require a more robust Bayesian sampling framework still capable of leveraging differentiable models, such as \code{Comrade.jl} \citep{Tiede_2022}.

Future work will focus on extending \code{Jipole}’s capabilities in several directions. First, we plan to incorporate GRMHD simulation data as input, together with a thermal synchrotron emission model \citep{Leung_2011}, enabling the use of physically motivated plasma descriptions in place of simplified analytical prescriptions. Second, we aim to add polarization calculations, allowing for more comprehensive comparisons with polarimetric observations. Third, this formulation naturally extends to a slow-light ray-tracing mode, in which the finite speed of light is explicitly accounted for, thereby incorporating time delays and the temporal evolution of the emitting medium into the integration. Finally, we intend to implement analytical geodesic integration schemes to replace numerical integrators where feasible. This enhancement will be particularly important for polarization studies, since polarization transport requires propagating additional variables along the geodesics. Analytical integration will reduce the need to cache these variables, thereby improving both computational efficiency and memory usage.

\begin{acknowledgments}

We thank Dominic Chang, Nick Conroy, Charles Gammie, Trevor Gravely, Hyun Lim, Rodrigo Nemmen, Paul Tiede, George Wong and Aristomenis Yfantis for their helpful comments and suggestions. We also thank Donald Thompson for computing support. P. N. M. was supported by FAPESP (Fundação de Amparo à Pesquisa do Estado de São Paulo) under grant 2023/15835-2. P. N. M. and A. C.-A. acknowledge support from the Center for Nonlinear Studies at Los Alamos National Laboratory. A. C.-A. and B.S.P. were supported by the US Department of Energy through the Los Alamos National Laboratory. Los Alamos National Laboratory is operated by Triad National Security, LLC, for the National Nuclear Security Administration of U.S. Department of Energy (Contract No. 89233218CNA000001). This work has been assigned a document release number LA-UR-25-28992.

\end{acknowledgments}

\bibliography{citations}{}
\bibliographystyle{aasjournal}

\end{document}